\documentclass[twoside]{article}
\usepackage{fleqn,espcrc2}
\usepackage{amsmath}
\usepackage{amssymb}
\usepackage{graphicx}

\title{Ginzburg-Landau theory for the time-dependent phase field 
in a two-dimensional $d$-wave superconductor}

\author{S.G. Sharapov\address[Neuchatel]{Institut de Physique, 
Universite de Neuch\^atel, Neuchatel, Switzerland}%
        \thanks{On leave of absence from Bogolyubov Institute for 
         Theoretical Physics, 03143 Kiev, Ukraine},
        H. Beck\addressmark\thanks{{\tt E-mail:} Hans.Beck@unine.ch
         {\tt Fax:} 41 32 718 2901 },
        and V.M. Loktev\address{Bogolyubov Institute for Theoretical Physics, 
                03143 Kiev, Ukraine}}
       
\begin{document}

\begin{abstract}
We derive a finite temperature time-dependent effective theory
for the phase $\theta$ of the pairing field, which is appropriate 
for a 2D conducting electron system with non-retarded d-wave attraction. 
As for $s$-wave 
pairing the effective action contains terms with Landau damping,
but their structure appears to be different from the $s$-wave case 
due to the fact that the Landau damping is determined by the
quasiparticle group velocity ${\bf v}_{g}$, which for the $d$-wave
pairing does not have the same direction as the non-interacting 
Fermi velocity ${\bf v}_{F}$. We show that for the $d$-wave pairing the Landau
terms have a linear low temperature dependence and in contrast
to the $s$-wave case are important for all finite temperatures.\\
\vspace{1pc}
{\tt Key words:} time-dependent Ginzburg-Landau theory, $d$-wave pairing
\end{abstract}

\maketitle

1.~The microscopic derivation of the effective time-dependent Ginzburg-Landau
(GL) theory continues to attract attention since an early paper
by Abrahams and Tsuneto \cite{Abrahams}. Whereas the static GL potential
was derived 
from the microscopic BCS theory soon after its 
introduction, the time-dependent GL theory is still a subject of interest
\cite{Aitchison}.
One of the reasons for this is the presence of Landau damping terms
in the effective action. For $s$-wave superconductivity these terms
are singular at the origin of energy-momentum space, and consequently
they cannot be expanded as a Taylor series about the origin.
In other words, these terms do not have a well-defined expansion
in terms of space and time derivatives of the ordering field and therefore
they cannot be represented as a part of a local Lagrangian.
We recall that at $T =0$ and for the static, time-independent case
the Landau damping vanishes, so that either at $T =0$ one still has
a local well-defined time-dependent GL theory or for $T \neq 0$
the familiar static GL theory exists. It is known, however,
that for $s$-wave superconductivity even though the Landau terms
do exist, they appear to be small compared to the main terms of
the effective action in the large temperature region $0 < T \lesssim 0.6 T_c$
\cite{Aitchison}, where $T_c$ is the superconducting
transition temperature. 
This is evidently related to the fact that only
thermally excited quasiparticles contribute to the Landau damping.
The number of such quasiparticles at low temperatures is 
an exponentially small
fraction of the total charge carriers number in the $s$-wave
superconductor due to the nonzero superconducting gap
$\Delta_s$ which opens over all directions on the Fermi surface.

For a $d$-wave superconductor there are four points (nodes)
where the superconducting gap $\Delta_{d}({\bf k}) =0$ 
on the Fermi surface. 
The presence of the nodes increases significantly the number of the 
thermally excited quasiparticles at given temperature $T$ comparing to 
the $s$-wave case. Therefore one can expect that the Landau damping 
is stronger for superconductor with a 
$d$-wave gap which is commonly accepted to be the case of high-temperature 
superconductors (HTSC). Moreover, it is believed that
at temperatures $T \ll T_c$,these quasiparticles are reasonably
well described by the Landau quasiparticles, even though such an
approach fails in these materials at higher energies \cite{Lee}. 
This is the reason why one can hope that a generalization of the BCS-like
approach \cite{Aitchison} for the 2D $d$-wave superconductivity may be
relevant to the description of the low-temperature time-dependent
GL theory in HTSC.

We derive such a theory from a microscopical model with
$d$-wave pairing extending the approach of \cite{Aitchison} developed
for $s$-wave superconductivity. As known from \cite{Abrahams}
the physical origin of the Landau damping is a scattering of the
thermally excited quasiparticles (normal fluid) with the group
velocity $\mathbf{v}_g$ from the phase 
(or $\theta-$) excitations (quantums). Such 
scattering occurs only if the {\v C}erenkov 
irradiation (absorption) condition,
$\Omega = \mathbf{v}_g \mathbf{K}$ for the energy $\Omega$ and momentum
$\mathbf{K}$ of the $\theta$-excitation is satisfied. This phenomenon
in superconductivity is also called  Landau damping since its
equivalent for the plasma theory  was originally obtained by Landau.

One of the  main physical differences between the $s$-
and $d$-wave cases is related to the fact that for $d$-wave 
superconductivity the direction of the quasiparticle group
velocity $\mathbf{v}_{g} (\mathbf{k}) \equiv 
\partial E(\mathbf{k})/\partial \mathbf{k}$
($E(\mathbf{k})$ is the quasiparticle dispersion law) does not
coincide with the Fermi velocity $\mathbf{v}_F$ 
\cite{Lee,Carbotte} and a gap velocity  
$\mathbf{v}_{\Delta} \equiv \partial \Delta_{d}(\mathbf{k})/\partial \mathbf{k}$
also enters into the {\v C}erenkov condition along with $\mathbf{v}_F$.
We also show that the intensity of the Landau damping 
is proportional to $T$ at low temperatures. 

2.~We consider the Hamiltonian $H$ for fermions 
on the square lattice with the lattice constant $a$,
the dispersion law $\xi(\mathbf{k})$  and the attractive  
potential $V(\mathbf{r})$ the momentum representation of which
contains only $d$-wave pairing
\begin{equation}
\label{Hamilton}
\begin{split}
\! \! \! \! \!  
H  = &  \sum_{\sigma} \int d \tau \left\{ \int d^2 r 
\psi_{\sigma}^{\dagger}(\tau, {\bf r})
\xi(- i \nabla) \psi_{\sigma}(\tau, {\bf r})   \right.   \\
& -  \frac{1}{2}  \int d^2 r_1 \int d^2 r_2
\psi_{\sigma}^{\dagger}(\tau, {\bf r}_2) \psi_{\bar{\sigma}}^{\dagger}(\tau,
{\bf r}_1) \\
& \left. 
\times V({\bf r}_1; {\bf r}_2) \psi_{\bar{\sigma}}(\tau, {\bf r}_1)
\psi_{\sigma}(\tau, {\bf r}_2) \right\}\,.
\end{split}
\end{equation}
Here  $\psi_{\sigma}(\tau, {\bf r})$ is a fermion field with the spin 
$\sigma= \uparrow, \downarrow$, $\bar{\sigma} \equiv - \sigma$
and $\tau$ is the imaginary time.
The final results will be formulated in terms of 
the Fermi  $\mathbf{v}_F \equiv \partial \xi(\mathbf{k})/\partial \mathbf{k}
|_{\mathbf{k}= \mathbf{k}_F}$ and
the gap $\mathbf{v}_\Delta$ velocities which  proved to be very
convenient both in the theory of $d$-wave superconductors  
\cite{Lee,Carbotte} and for the analysis of various experiments \cite{Chiao}. 

The Hubbard-Stratonovich method is employed to derive the effective
``phase-only'' action (see the review \cite{Loktev.review} and Refs. therein).
The present derivation has some specific features related to
the non-local character of the interaction in coordinate
space, so that a bilocal Hubbard-Stratonovich field has to be used 
\cite{Kleinert} and an additional Born-Oppenheimer approximation is necessary 
to separate the terms describing a relevant phase dynamics from
the rest of the effective action. The detail of this rather lengthy
calculation will be presented elsewhere \cite{we}, so that here
we present only the final results.

\noindent
3. If the Landau terms are neglected it is possible to express the 
thermodynamical potential $\beta \Omega_{\rm kin} = - i \int dt \int d^2 r 
\mathcal{L}^{\mbox{\tiny R}}(t, \mathbf{r})$ ($\beta \equiv 1/T$)
in terms of a local effective Lagrangian
\begin{equation}
\label{Lagrangian}
\mathcal{L}^{\mbox{\tiny R}} = - 
\frac{n_f}{2} \dot{\theta}(t, {\bf r}) + \frac{K}{2}
 [\dot{\theta}(t, {\bf r})]^2 - 
\frac{J}{2}[\nabla \theta (t, {\bf r})]^2\,,
\end{equation}
which is valid for $T \ll \Delta_{d}$, where $\Delta_d$
is the amplitude of the superconducting gap and $n_f$ is the carrier
density. In (\ref{Lagrangian})
\begin{equation}
\label{kinetic.d-wave}
J = \left(\frac{\sqrt{\pi v_{F} v_{\Delta}}}{24 a} - \frac{\ln 2}{2 \pi}
\frac{v_F}{v_{\Delta}}T\right),   
K = \frac{1}{4 a \sqrt{\pi v_{F} v_{\Delta}}} 
\end{equation}  
are the phase stiffness and compressibility, respectively.
The Lagrangian (\ref{Lagrangian}) describes the collective phase
excitations (Berezinskii-Kostrelitz-Thouless mode) which is the 2D
analog of the well-known 3D Bogolyubov-Anderson mode. The Landau
terms which will be considered in what follows are, in fact,
the corrections to (\ref{Lagrangian}) non-local in coordinate, 
which result in the damping of $\theta$-excitations.

\noindent
4. The Landau terms originate from the terms 
\begin{equation}
\sim \frac{(\mathbf{v}_{F} \mathbf{K})^{\alpha}} 
{\mathbf{v}_{g} \mathbf{K} - \Omega - i0} \frac{d n_{F} (E)}{d E}  
\mathbf{v}_{g} \mathbf{K} \quad (\alpha = 0,1,2)
\end{equation}
which are present in the momentum - real frequency representation
of the effective action \cite{we} ($n_{F}$ is the Fermi distribution). 
The proper treatment of these denominators leads to the following result:
\begin{equation}
\label{effective.action.Landau}
\begin{split}
\beta \Omega_{kin}^{\tiny L} = & 
\frac{i}{2}  \int d \Omega \int \frac{K dK}{2 \pi}
\int_{0}^{2 \pi} \frac{d \phi}{2 \pi} \\
& \theta(\Omega, \mathbf{K})
F_{s,d}^{\mbox{\tiny L}}(\Omega, K, \phi) \theta(-\Omega, -\mathbf{K})\,.
\end{split}
\end{equation}
For $s$-wave pairing \cite{Aitchison} all three terms with $\alpha =0,1,2$
result in $F_s^{\mbox{\tiny L}} \sim \Omega^3/v_{F} K$ 
because $\mathbf{v}_{g} \parallel \mathbf{v}_F$. In this case 
the intensity of the Landau damping does not depend on the direction of 
the vector $\mathbf{K} = (K \cos(\phi - \pi/4), K \sin (\phi - \pi/4))$ 
in the plane. (The angle $\phi$ is chosen in such a way that $\phi = \pi/4$
corresponds to the first node of the $d$-wave gap when $d$-wave
pairing is considered.) Furthermore, the damping process is 
exponentially suppressed by the factor $\exp(-\Delta_s/T)$ 
($\Delta_s$ is the $s$-wave superconducting gap)
reflecting a small number of thermally excited quasiparticles
at $T \ll \Delta_s$ which contribute into the Landau damping.

\begin{figure}[h!]
\vspace{-0.6cm}
\centering{
\includegraphics[width=6.5cm]{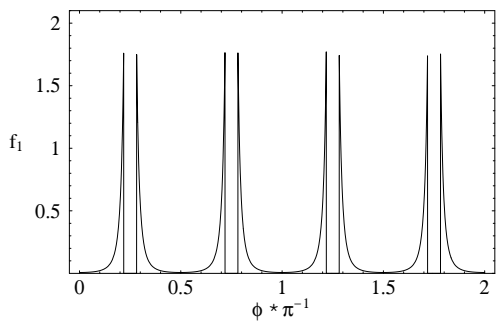}}
\vspace{-1cm}
\caption{The angular dependence, 
$f_1 \left(\phi \right)$
of the Landau term $\sim \Omega^3/K$ for $v_F / v_{\Delta} =20$
and $|\Omega|/v_F K =0.1$.}  
\vspace{-0.9cm}
\label{fig:1}
\end{figure}

Since for $d$-wave pairing $\mathbf{v}_{g} \nparallel \mathbf{v}_{F}$,
the terms with different $\alpha$ do not produce the same analytical
structure with all terms $\sim \Omega/v_f K$ as for $s$-wave
pairing. As the result 
we arrive at the following more complex expression:
\begin{equation}
\label{effective.action.singular}
\begin{split}
&  F_{d}^{\mbox{\tiny L}}   (\Omega, K, \phi) = -i 
\frac{T \ln 2 }{v_{\Delta} 4\pi}\left(  
\frac{\Omega^3}{K} 
\frac{1}{v_{F}^{2}} f_1 \left( \phi \right) + \right.  \\
& \left.  \Omega K   
f_2 \left(\phi \right) - \Omega^2
\frac{2}{v_F} f_3 \left( \phi \right) \right) \,, \quad 
\frac{\Omega}{v_{F} K} \ll 1\,,
\end{split}
\end{equation}
where the functions $f_{1}$, $f_2$ and $f_3$ obtained in \cite{we} describe
the directional dependence of the corresponding damping term. 
The analytical expression for these functions are simple, but rather
lengthy, so that here we show only the graphic for one them
in Fig.~\ref{fig:1}.
It appears that the calculation of ultrasonic attenuation
for $d$-wave superconductivity \cite{Carbotte} gives a result
similar to the first term of (\ref{effective.action.singular})
and indeed the angular dependence described by $f_1$ coincides
with the dependence obtained in \cite{Carbotte}. The presence
of angular dependence in (\ref{effective.action.singular})
demonstrates explicitly that the intensity of damping depends
on the direction of $\theta$-particle motion with respect to the nodes
on the Fermi surface.  One can see from (\ref{effective.action.singular})
that the Landau damping is linear in $T$ and thus it is much
stronger than for the $s$-wave case.

\noindent 
5. It is very important to recall that the collective phase excitations
described here can and have been studied experimentally. Indeed 
the measurements of the order parameter dynamical structure factor in the 
dirty Al films allowed to extract the dispersion relation of the 
corresponding Carlson-Goldman
mode and to investigate its temperature dependence 
\cite{Goldman}. The model considered here shows that
it would be interesting to address experimentally the physics of the 
phase excitations in $d$-wave superconductors \cite{Ohashi} which as 
we have demonstrated has many specific features.

This work was  supported by the  research project 2000-061901.00/1
and SCOPES-project 7UKPJ062150.00/1
of the Swiss National Science Foundation. The work of V.M.L. is partially 
supported by NATO grant CP/UN/19/C/2000/PO.

\end{document}